\documentclass[twocolumn,prl,superscriptaddress]{revtex4-2}
\usepackage{bm}
\usepackage[fleqn]{amsmath}
\usepackage{hyperref}
\usepackage{graphicx}
\usepackage{lipsum}
\usepackage{siunitx}

\begin{document}

\title{Non-chiral one-dimensional states propagating inside AB/BA domain walls in bilayer graphene }
\author{V.V. Enaldiev}
\author{C. Moulsdale}
\address{National Graphene Institute, University of Manchester, Booth St. E. Manchester, M13 9PL, United Kingdom}
\address{Department of Physics and Astronomy, University of Manchester, Oxford Road, Manchester, M13 9PL, United Kingdom}
\author{A.K. Geim}
\address{National Graphene Institute, University of Manchester, Booth St. E. Manchester, M13 9PL, United Kingdom}
\address{Department of Physics and Astronomy, University of Manchester, Oxford Road, Manchester, M13 9PL, United Kingdom}
\author{V.I. Fal'ko}
\email{vladimir.falko@manchester.ac.uk}
\address{National Graphene Institute, University of Manchester, Booth St. E. Manchester, M13 9PL, United Kingdom}
\address{Department of Physics and Astronomy, University of Manchester, Oxford Road, Manchester, M13 9PL, United Kingdom}
\address{Henry Royce Institute for Advanced Materials, University of Manchester, Manchester, M13 9PL, United Kingdom}

\date{\today}

\begin{abstract}
Boundaries between structural twins of bilayer graphene (so-called AB/BA domain walls) are often discussed in terms of the formation of topologically protected valley-polarised chiral states. Here, we show that, depending on the width of the AB/BA boundary, the latter can also support non-chiral one-dimensional (1D) states that are confined to the domain wall at low energies and take the form of quasi-bound states at higher energies, where the 1D bands cross into the two-dimensional spectral continuum. We present the results of modeling of electronic properties of AB/BA domain walls with and without magnetic field as a function of their width and interlayer bias.      
\end{abstract}

\maketitle

Bilayer graphene (BLG) has emerged as a versatile material system with electronic properties tunable by an electric field \cite{Oostinga2007,MorpurgoPRL2008,Zhang2009,Varlet2014}, strain and the interlayer twist angle \cite{cao2018correlated,cao2018unconventional,Rickhaus2018,Yankowitz2019,Xu2019,yoo2019atomic,lu2019superconductors,sharpe2019emergent,cao2021nematicity}. In bilayers with a standard Bernal stacking, the out-of-plane electric field opens an interlayer-asymmetry spectral gap, transforming a gapless to a gapfull semiconductor \cite{PhysRevLett.96.086805,McCann2006,Oostinga2007,Zhang2009,Slizovskiy2021}. Moreover, a BLG with an interlayer-asymmetry gap has been identified as 'a weak topological insulator' characterised by valley Chern numbers $\pm2$ \cite{ZhangMacDMele2013}, which manifests itself in chiral valley-polarized gapless modes confined around the gap inversion line \cite{MorpurgoPRL2008}. 

In terms of the lattice structure, Bernal bilayers can exist in the form of two twins known in the graphite literature as AB and BA stacking configurations \cite{Delavignette1960}, both of which may be realized in a single BLG flake as mesoscale domains. AB/BA boundaries separating pairs of such twins have been identified in several STEM and STM studies of both exfoliated monolithic bilayers \cite{Butz2013} and marginally twisted ones (assembled from two monolayers by the orientation-controlled flake transfer) \cite{AldenPNAS,yoo2019atomic}. This AB/BA boundaries are nothing but partial dislocations, which can be distinguished from perfect ``full'' dislocations (separating identical stacking domains, AB/AB or BA/BA) by their Burgers vector \cite{hirth1992,Kosevich2005}. Due to a weak topological nature of gapped BLG, these partial dislocations host valley-polarized chiral modes inside the interlayer-asymmetry gap of the bilayer \cite{ZhangMacDMele2013,Ju2015,Yin2016}. For a given domain boundary the direction of propagation of chiral states is set by the valley index of electrons (reversed in opposite valleys). In marginally twisted bilayers this gives rise to triangular networks of valley-polarized chiral ballistic conductors with suppressed electron backscattering at the network nodes \cite{SanJose2013,Efimkin2018,Huang2018} manifested in magnetotransport measurements by characteristic Aharonov-Bohm oscillations \cite{Rickhaus2018,Xu2019,Beule2021}.

\begin{figure}
    \centering
    \includegraphics[width=\columnwidth]{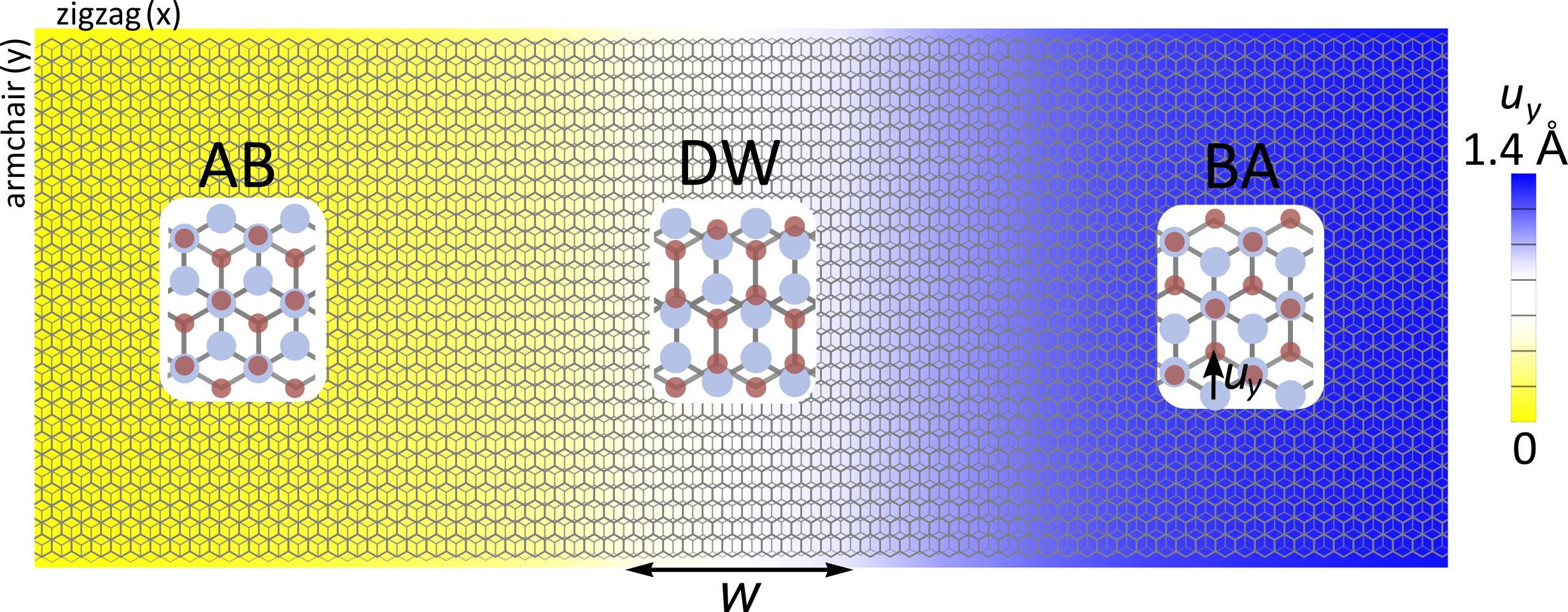}
    \caption{AB/BA domain wall in bilayer graphene, representing partial screw dislocation that is characterised by shear displacements $u_y(x)$ along armchair direction. Insets show stackings in domains and in the middle of domain wall. \label{Fig:1}}
\end{figure}

In addition to valley-polarized chiral states, other electronic states may be localised within a domain boundary. For example, the abundance of such states was hinted by the studies of domain walls (DWs), modeled as an interval of two separated monolayers (a delamination) \cite{Lane2018}, or as largely extended crossover region between two Bernal stacking areas \cite{Shallcross2022}. In contrast to the chiral states, electrons occupying these additional states can propagate in both direction in each valley, and, therefore, undergo back-scattering within each stretch of domain boundary, thus qualitatively changing properties of the whole DW network. Here, we study confinement of states at a single AB/BA domain boundary in BLG, both gapless and with an inter-layer potential, at zero and finite magnetic field. For a gapless bilayer we notice the appearance of non-chiral 1D bands localised at the AB/BA boundary when their energies are outside the bulk BLG dispersion. These non-chiral 1D bands can also extend into the BLG spectral continuum, where they become quasistationary due to escaping into the surrounding AB and BA domains. Nevertheless we find a way to trace those by following sharp dispersive features in the Landau level spectrum computed for the same system, where the identity of localised bands is partly restored by the discreteness of the Landau level spectrum.  In gapped BLG these non-chiral states can coexist with the valley-polarized chiral modes, however, with the spectra dependent on the inter-layer potential and sensitive to the DW width, so that below we explore an experimentally  relevant parameter space of the AB/BA DW band structure.

\section{Frenkel-Kontorova model for an AB/BA domain boundary}

To describe the lattice structure of a single AB/BA DW in BLG we adopted the Frenkel-Kontorova model (FKM) for a partial dislocation which is a lateral boundary between two twin crystals. Such a DW is characterized by Burgers vector $\bm{b}=(0,a/\sqrt{3})$ aligned with the armchair direction that sets boundary conditions for an interlayer displacement vector $\bm{u}=(0,u_y(x))$ describing a transformation of an AB into BA stacking, Fig. \ref{Fig:1}. The armchair orientation of the DW axis is also known to be energetically favourable \cite{AldenPNAS,Zhou2015,yoo2019atomic,Enaldiev_PRL} (i.e.   this is true partial screw dislocation for which Burgers vector is parallel to its axis), the minimal FKM can be formulated in terms of a single component interlayer displacement field $u_y(x)$ and its solution minimises the sum of elastic and adhesion energies \cite{Kosevich2005,Popov2011,AldenPNAS,Lebedeva2016,Lebedeva2019},
$$ {\cal U}=\int_{-\infty}^{\infty}\left\{\frac{\mu}{4}\left(\frac{\partial u_y}{\partial x}\right)^2+{\cal V}\left[1-\cos\left(2\pi\frac{u_y\sqrt{3}}{a}\right)\right]\right\}dx.$$
Here, the coordinate $x$ accounts for DW cross section along zigzag direction, $a$ and $\mu$ are lattice constant and shear modulus of monolayer graphene, respectively, and parameter ${\cal V}$ describes adhesion energy landscape for a pair of 2D hexagonal crystals \cite{Popov2011,Zhou2015,CarrPRB2018,Enaldiev_PRL}. We note that the actual value of ${\cal V}$ for a given structure may depend on whether one considers free-standing or encapsulated bilayer and also may be weakly temperature-dependent. The above energy functional is minimised by the following interlayer displacement field profile \cite{Kosevich2005,Popov2011,AldenPNAS,Lebedeva2016,Lebedeva2019},
\begin{equation}
\label{Eq:dw}
    u_y(x)=\frac{2a}{\sqrt{3}\pi}\arctan\left(e^{2x/w}\right),
\end{equation}   
where $w=(a/\pi)\sqrt{\mu/6{\cal V}}$ characterises the full width of the DW \footnote{Substituting in Eq. \eqref{Eq:dw} $a=0.246$\,nm,  $\mu= 920$\,eV/nm$^2$\, and  $V=0.0296$\,eV/nm$^2$ \cite{CarrPRB2018,Zhou2015} one obtains $w\approx 6$\,nm.}. Despite several studies of AB/BA domain walls, there is no a commonly-established value for the DW width, likely due to strain induced by the flake transfer and variation of encapsulation conditions.  The reported/discussed values range as $6$\,nm $<w<$ $14$\,nm \cite{AldenPNAS,Yin2016,Lebedeva2016}. Therefore, in the following studies we implement $w$ as a phenomenological parameter and, when modeling electronic properties of an AB/BA DW, cover the above range of $w$ values. 

\section{DW-bound electron states}

The quantitative description of electron states in a bilayer with variable stacking $\bm{r}_0(x)$ (defined to produce AA-stacking for $\bm{r}_0=0$) is obtained using a 4x4 Hamiltonian \cite{PhysRevB.104.085402}:
\begin{eqnarray}\label{Eq:init_H}
    \mathcal{H}_{\xi}(\bm{r}_0)=
    \begin{pmatrix}
        v\bm{\sigma}_{\xi}\bm{p}+\frac{\Delta}{2} & T_{\xi}(\bm{r}_0)\\
        T^{\dag}_{\xi}(\bm{r}_0) & v\bm{\sigma}_{\xi}\bm{p} -\frac{\Delta}{2}
    \end{pmatrix}, \qquad\quad\\
     T_{\xi}(\bm{r}_0)= \frac{\gamma_1}{3}\sum_{j=0,1,2}e^{i\xi(\bm{K}_j-\bm{K}_0)\bm{r}_0}\begin{pmatrix}
        1 & e^{i\xi2\pi j/3}\\
        e^{-i\xi2\pi j/3} & 1 
        \end{pmatrix}, \nonumber \\
    \bm{r}_0(x)=\left(0,\frac{a}{\sqrt{3}}+u_y(x)\right). \qquad\qquad\qquad\nonumber
\end{eqnarray}
Here, we use a four-component wave function $\Psi(\bm{r})=(\psi^{t}_{A},\psi^t_{B},\psi^b_{A},\psi_{B}^b)^{\rm T}$, written in layer and sublattice basis; $v\approx10^6$\,m/s is Dirac velocity in monolayer graphene, $\bm{p}$ is valley-momentum counted from the $\pm{\rm K}$-point, labeled by $\xi=\pm$, $\sigma_{\xi}=(\xi\sigma_x,\sigma_y)$ is a vector of Pauli matrices acting on sublattice components in each layer, and $\Delta$ is interlayer potential difference. In the tunneling matrix $T_{\xi}$, $\gamma_1\approx 380\,$meV and \mbox{$\pm\bm{K}_{j=0,1,2}=\pm(4\pi/3a)\left[\cos(2\pi j/3),-\sin(2\pi j/3)\right]$} are two tri-stars of equivalent Brillouin zone corners corresponding to $\pm$K-valleys. For $\bm{r}_0=(0,a/\sqrt{3})$ and $\bm{r}_0=(0,2a/\sqrt{3})$ the interlayer coupling $T_{\xi}$ reproduces the minimal Hamiltonians for AB and BA Bernal BLG, respectively, whereas in the inner part of DW it can be represented as 
\begin{align}\label{Eq:T_ABBA}
    T_{\xi}(x)&=\begin{pmatrix}
        t_0(x) & t_1(x) \\
        t_1(x)+2 t_2(x) & t_0(x)
    \end{pmatrix},
\end{align}
\begin{align}
    t_0(x)&=\frac{\gamma_1}{3}\left\{1+2\cos\left(\frac{2\pi}{3}\left[\frac{u_y(x)\sqrt{3}}{a}+1\right]\right)\right\},\nonumber\\
    t_1(x)&=\frac{\gamma_1}{3}\left\{1+2\cos\left(\frac{2\pi}{3}\frac{u_y(x)\sqrt{3}}{a}\right)\right\},\nonumber\\
    t_2(x)&=\frac{\gamma_1}{\sqrt{3}}\sin\left(\frac{2\pi}{3}\left[\frac{u_y(x)\sqrt{3}}{a}-\frac12\right]\right)\nonumber.
\end{align}

\begin{figure*}
    \centering
    \includegraphics[width=1.5\columnwidth]{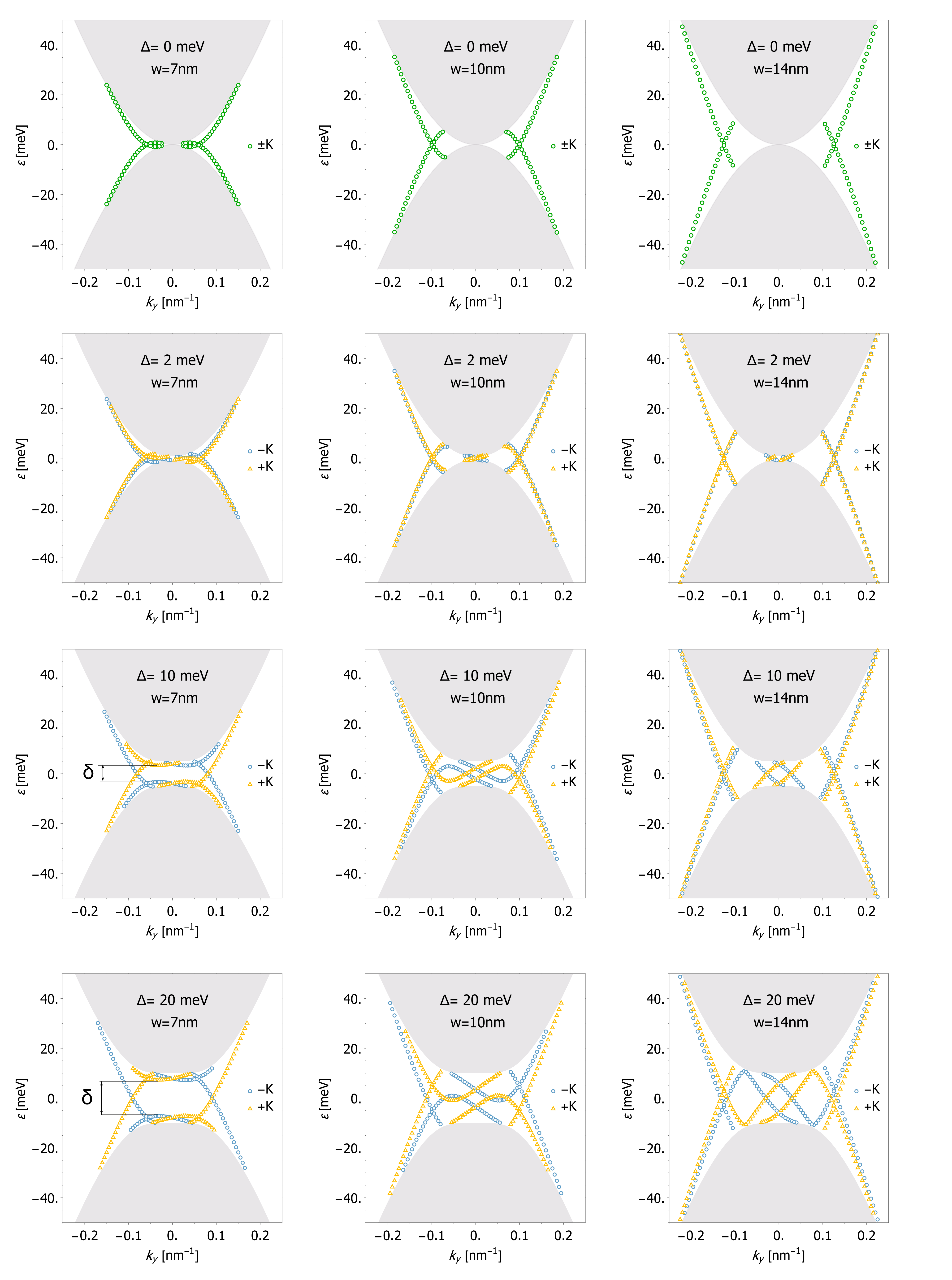}
    \caption{Dots show computed dispersion of DW-localised 1D bands shown in $\pm$K valleys for three values of DW widths, $w=7,10,14$\,nm and interlayer-asymmetry potential $\Delta=0, 2, 10, 20$\,meV labeled on each panel. Gray regions display continuum of scattering states. For $\Delta=0$ (top row) all states are double degenerate on valley index $\xi=\pm$ due to $\hat{\mathcal{C}}$ and $\hat{\mathcal{S}}$ symmetries, determined by Eqs. \eqref{Eq:charge_symm} and \eqref{Eq:str_symm}, respectively. Minigap, $\delta$, in spectrum of the non-chiral states is shown in the left column panels.}
    \label{fig:spectra_B=0}
\end{figure*}

Using translation-invariance along the DW axis we describe all states as plane waves in $y$ direction characterised by longitudinal momentum $p_y$. For each $p_y$ we diagonalise Hamiltonian \eqref{Eq:init_H} numerically by projecting the problem on a coordinate grid $x_{n}=n s-L/2$ set across the interval $-L/2\leq x\leq L/2$ with a step $s\ll w$. This discretisation is realized by replacing $\hat{p}_x\Psi$ with $-i\hbar(\Psi_{n+1}-\Psi_{n-1})/2s$ (where $\Psi_{n}=\Psi(x_n)$) and solving a homogeneous system $4L/s$ of simultaneous equations,
\begin{equation}\label{Eq:fin_dif_sys}
    \frac{i\xi}{2s}\sigma_x\left[\Psi_{n-1}-\Psi_{n+1}\right]+
    \left[\mathcal{H}_{\xi}(0,p_y,x_n)-\varepsilon\right]\Psi_{n}=0,
\end{equation}
with open boundary conditions. Here, each $4\times4$ matrix $\mathcal{H}_{\xi}(0,p_y,x_n)$ is defined by replacing $p_x\to0$ in Eq. \eqref{Eq:init_H}. 

The computed energy spectra (which convergence tested by using both a finer grid and a larger system size $L$) are shown in Fig. \ref{fig:spectra_B=0}. They include two types of states:
\begin{itemize}
    \item a continuum of scattering states, spread over the AB and BA areas covering energy range, $\varepsilon^2\geq \frac{\Delta^2}{4}+\frac{\gamma_1^2}{2}+(vp_y)^2-\sqrt{\frac{\gamma_1^4}{4}+\left(\Delta^2 +\gamma_1^2\right) (vp_y)^2}$,
    \item states localised at DW with energies outside the continuum. 
\end{itemize}
For a zero inter-layer potential, $\Delta=0$, the computed spectra are both electron-hole symmetric and valley degenerate, see top row in Fig. \ref{fig:spectra_B=0}. This symmetry of Hamiltonian \eqref{Eq:init_H}, which, for $\Delta=0$, features both charge conjugation symmetry, 
\begin{equation}\label{Eq:charge_symm}
   \hat{\mathcal{C}}\mathcal{H}_{\xi}\hat{\mathcal{C}}^{+}=-\mathcal{H}_{\xi},\quad \hat{\mathcal{C}}=\tau_z\sigma_x\hat{M}_{x},
\end{equation} 
where $\hat{M}_x$ is a $x\to-x$ reflection, and 
\begin{equation}\label{Eq:str_symm}
\hat{\mathcal{S}}\mathcal{H}_{\xi}(p_x,p_y)\hat{\mathcal{S}}^{+}=\mathcal{H}_{\xi}(p_x,-p_y), \quad \hat{\mathcal{S}}=\tau_x\sigma_x.
\end{equation}
In the above transformations $\tau_{x,y,z}$ are Pauli matrix acting in the layer space. While the first of these symmetry relations determines the electron-hole symmetry of the spectrum, the second leads to a $\varepsilon(p_y)=\varepsilon(-p_y)$ symmetry in both valleys. As a result the identified localised states support counter-propagating modes in both $+$K and $-$K valleys, hence, they are non-chiral.

Breaking the bilayer inversion symmetry by interlayer potential bias, $\Delta$, qualitatively changes the bilayer spectra. The resulting spectra no more have the above-described electron-hole and $\varepsilon(p_y)=\varepsilon(-p_y)$ symmetries, however, they reflect time-reversal symmetry of the model by reverted $p_y$-dispersion in the opposite valleys. First of all, it generates pairs of valley-polarised chiral states inside the gap. Increasing $\Delta$ also modifies the dispersion of the non-chiral 1D bands: as shown in Fig. \ref{fig:spectra_B=0}, the chiral and non-chiral bands merge and reconnect their dispersion lines, which happens at higher $\Delta$s in wider DWs. Also, upon increasing $\Delta$, the edges of the reconnected bands are pushed apart towards $\pm \Delta/2$ band edges in bulk BLG, crossing over from the overlapping bands regime at intermediate $\Delta$ values to a 'gapfull' pair at higher $\Delta$'s.

\section {DW in a magnetic field and manifestation of DW states {\it via} their hybridization with Landau levels}
\label{sec:magnetic}

When reaching the BLG continuum spectrum, Fig. \ref{fig:spectra_B=0}, the 1D bands of DW localised states become quasi-stationary due to a decay into the scattering states in the AB and BA stacking areas. In order to trace these 1D bands into these quasi-stationary regime we implement the following procedure. As an out-of-plane magnetic field transforms the 2D spectral continuum into discrete degenerate set of Landau levels, it would stabilize the 1D bands at the energies between bulk Landau levels. Then, instead of blending into a continuum spectrum of an infinite bulk, the 1D bands undergo a perturbative mixing with the Landau level states, forming together states propagating along the DW, manifested as a sharp bending of Landau level's dispersion. 

\begin{figure}
\centering
\includegraphics[width=\linewidth]{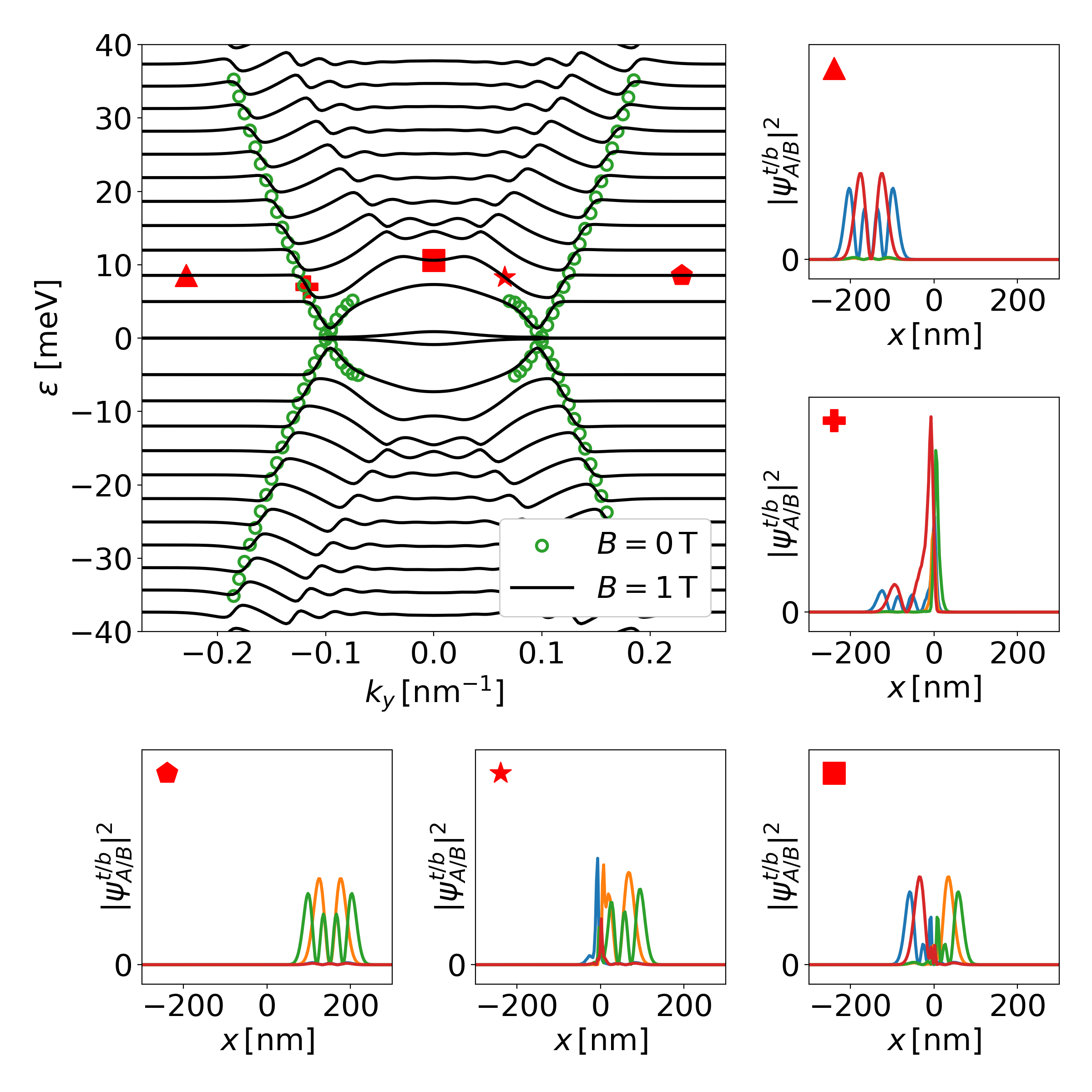}
\caption{\textit{Top left.} Dispersion of electron states in BLG with a single AB/BA domain wall in magnetic field $B=1$\,T as a function of conserving wave number $k_y$ for $w=10$\,nm and $\Delta=0$. \textit{Clockwise from top right.} The evolution of the spatial distribution of the electron density, $ |\psi | ^ 2 $ from Eq.~(\ref{eq:psib}), projected onto each of the sublattices, $ A_t$, $ B_t $, $ A_b $ and $ B_b$ (green, red, blue and amber lines, respectively), of a bulk AB state into that of a bulk BA state. Each panel corresponds to a wave function of a state labeled by corresponding symbol in the top left panel.}
\label{fig:density}
\end{figure}

The magnetic field effect is incorporated in Hamiltonian \eqref{Eq:init_H} via Peierls substitution \mbox{$\bm {p} \rightarrow ( -i\hbar\partial_x, - i\hbar\partial_y-eBx)$}. Then, we use a basis of magnetic oscillator states adapted to the Landau gauge to represent sublattice components of BLG states,  
\begin{equation}
\label{eq:psib}
\psi^{t/b}_{A/B} = e ^ {i k_y y}\sum_{n<N}\frac {C^{t/b}_{A/B,n}}{\sqrt {n!2 ^ n \sqrt\pi \lambda_B}}  e ^ {- z ^ 2/2} H_n (z),
\end{equation}
where $ z = x/\lambda_B - k_y \lambda_B $, and $ H_n $ are Hermite polynomials. The resulting matrix Hamiltonian is diagonalised numerically for a dense set of $k_y$ values, with the size $N$ of the basis choosing sufficiently large to guarantee convergence of the computed spectra in the range of $-100\,{\rm meV}\leq \varepsilon \leq 100\, {\rm meV}$. 

A typical spectrum computed for a BLG with an AB/BA boundary and $\Delta=0$ at $B=1\,$T is illustrated in Fig. \ref{fig:density}. For large values of $k_y$ it contains a ladder of non-dispersive Landau levels (LL) characteristic for a BLG \cite{PhysRevLett.96.086805}: these are the LL states inside the domains at the distances $|x|\sim |k_y|\lambda_b^2\gg w$, as exemplified  by the computed coordinate dependence of the eigenstates labeled by a triangle and a pentagon on the dispersion curves. In contrast to those dispersionless states and states marked by a square and a star, the eigenstate chosen at the dispersion point with the highest drift velocity (red cross in Fig. \ref{fig:density}) is strongly localised at the DW.

\begin{figure*}
\centering
\includegraphics[width=1.5\columnwidth]{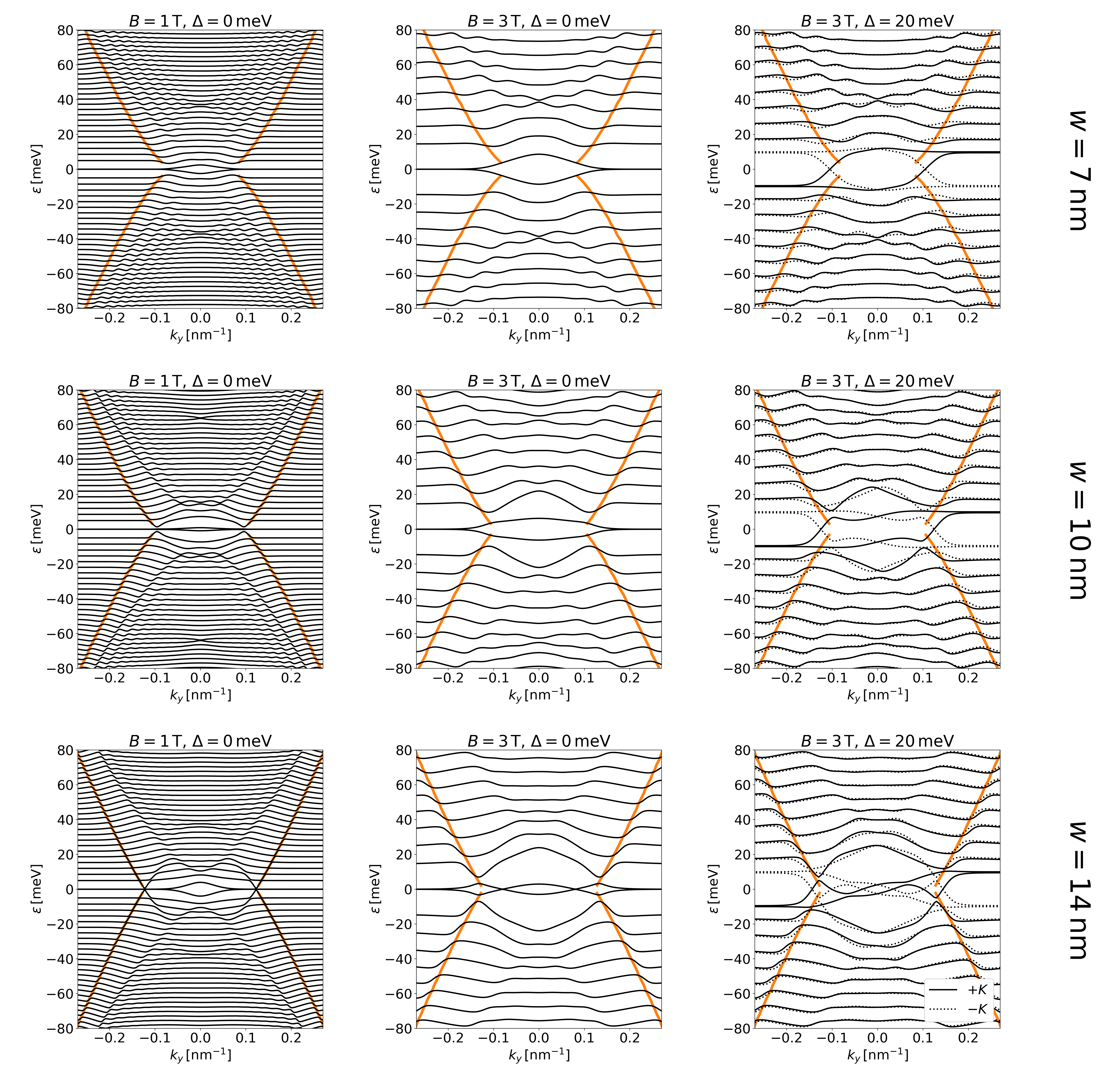}
\caption{Spectra of electron states in BLG with a single AB/BA domain wall (DW) for various magnetic fields, $B=1,\, 3$\,T, DW widths, $w =7, 10, \SI {14} {nm}$, and interlayer-asymmetry potentials, $\Delta=0,\, 20 $\,meV. For zero asymmetry, $\Delta=0$ in the two leftmost columns, the spectra are degenerate in the $\pm K$-valleys, which is lifted with non-zero asymmetry, $\Delta=20$\,meV in the right column. Orange lines, obtained by interpolation of the largest drift velocity states at $B = 1$\,T (left column) for each row (DW width), demonstrate dispersions of the non-chiral states. The dispersions, overlaid over spectra at $B=3\,$T and $\Delta=0, 20\,$meV in the middle and right columns, show that the largest-$ k_y $ kinks in spectra result from the non-chiral states.}
\label{fig:e}
\end{figure*}

When such fastest drifting states are traced across the spectra at $B=1$\,T for various DW widths, their energies fall onto $\varepsilon(k_y)$-dispersion of the localised 1D bands found at $B=0$ (replicated from Fig. \ref{fig:spectra_B=0} into Fig.  \ref{fig:e} as black dots). By following these spectral features beyond the spectral range of localised states identified in Fig. \ref{fig:spectra_B=0}, we are able to reconstruct the dispersion of quasi-bound DW states in the energy range falling inside the BLG spectral continuum at $B=0$, here shown by orange lines. An additional argument supporting this assignment comes from comparison of the spectra computed for $B=1$\,T and $B=3$\,T: for both of these fields the kinks in the dispersion correspond to the same $\varepsilon(k_y)$-dependence. This is in contrast to other dispersive features at higher energies, which positions scale up with magnetic fields and, therefore, can be attributed to the skipping cyclotron orbits of electrons moving in AB and BA domains within the cyclotron radius distance from the DW. 

While the DW spectra computed for $\Delta=0$ remain valley degenerate even in the presence of a magnetic field, the finite interlayer-asymmetry gap lifts such a degeneracy both inside the bulk areas \cite{PhysRevLett.96.086805} and on the DW itself. In the spectra shown on rhs panels in Fig. \ref{fig:e} one can identify the topological DW states that connect the sublattice-polarized `zero'-energy Landau levels in BLG. At the same time one can see that at the energies $|\varepsilon|\gg \Delta$ the dispersive features attributed to the LL mixing to the quasistationary 1D states remain almost non-effected. Overall, now, we are able to combine the dispersion of localised states computed for various DW widths at $B=0$ and extrapolated using the kinks in LL spectra to describe both truly localised and quasistationary states confined to the DW across a broad range of energies.

\begin{figure}
\centering
\includegraphics[width=\columnwidth]{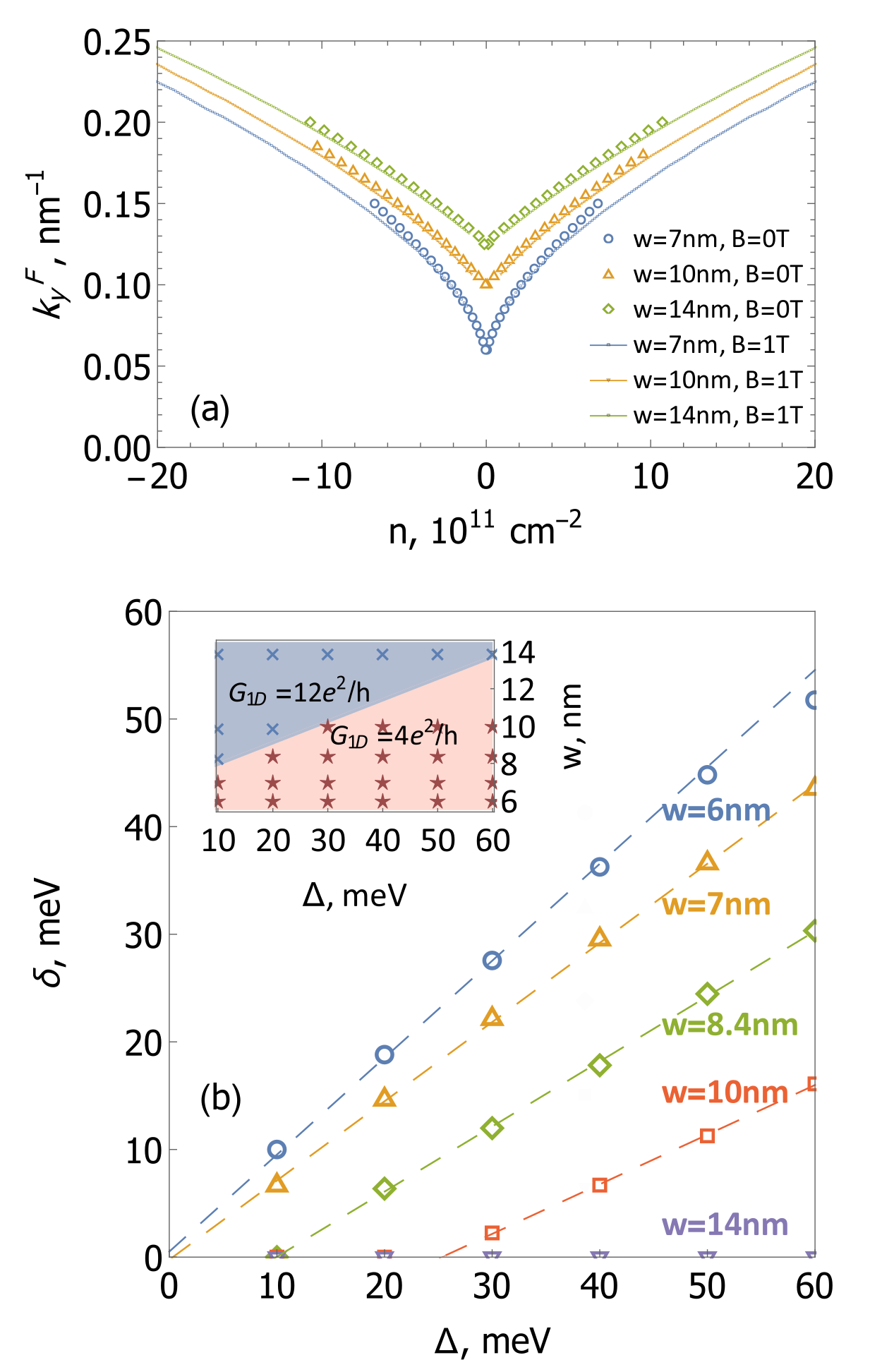}
\caption{(a) Dependence of Fermi momentum in non-chiral 1D band as a function of carrier concentration in BLG bulk. Dots correspond to dispersions at $B=0$, lines are calculated based on interpolated dispersions shown by orange in Fig. \ref{fig:e}. (b) Dependence of minigap (see left column in Fig. \ref{fig:spectra_B=0}) in non-chiral states spectrum as a function of interlayer-asymmetry potential. Inset: Blue and red regions characterise gapless and gapped spectra of the non-chiral states in $(\Delta,w)$ parametric space, which are characterised by difference of 1D conductance along DW at $\varepsilon_F=0$: $G_{1D}=12e^2/h$ for blue region and $G_{1D}=4e^2/h$ for red region, as explained in text. }
\label{fig:5}
\end{figure}

\section {Discussions \& Conclusions}
\label{sec:conclusion}

Based on the above-presented spectral data on the DW-confined non-chiral states we can discuss the experimental implications of their existence. Most strikingly, by inspecting the numerical data, we notice that such localised states coexist with a continuum spectrum even in a gapless BLG over a finite range of energies, $|\varepsilon| \leq 0.016 \gamma_1^2 w/\hbar v$, around neutrality point. This interval increases as $3.6\,[{\rm meV/nm}]\,\times w [{\rm nm}]$ with DW width. It is interesting to note that these 1D state dispersion crosses between conduction and valence band energy range at a finite value of $k_y$. This means that, while the electron Fermi momentum in the BLG bulk would scale down to zero with decreasing carrier density, the Fermi momentum in the 1D channel, that we estimate from the numerical data as $k_y^F\sim 0.03 w \gamma_1^2/(\hbar v)^2$, would remain finite at the neutrality point, as shown in Fig. \ref{fig:5}(a). We also note that, despite a decay of 1D channel states into 2D bulk at higher energies (or due to scattering of low energy 1D states by a residual disorder), in a segment of DW with a finite length they could act as a 1D quantum wire connected in parallel to the 2D bulk. For a BLG at a neutrality such a `wire' would contribute $8e^2/h$ (accounting spin and valley degeneracy) towards the two terminal conductance of a BLG device. 

Opening an interlayer-asymmetry gap both generates the valley-polarized chiral gapless states and modifies the dispersion of the non-chiral DW bands. Depending on the DW width and the size of interlayer-asymmetry gap $\Delta$ (see Fig. \ref{fig:spectra_B=0}) the spectrum of non-chiral states may remain gapless, or acquire a minigap, $\delta$. In Fig. \ref{fig:5}(b) we plot the computed values of the minigap $\delta$ as a function of input interlayer-asymmetry potential, $\Delta$, controllable by the out-of-plane displacement field \cite{Slizovskiy2021} over an experimentally feasible range. The computed values of $\delta$ represented by empty symbols and dashed lines are a linear fit drawn for each value of $w$ through the set of non-zero $\delta$ values. For each of the sampled DW widths $w$, the offset value in the linearly-extrapolated dependence $\delta(\Delta)$ was used to establish the number of 1D channels that would determine the conductance $G_{1D}$ of a BLG device when the chemical potential of electrons set right in the middle of the BLG spectral gap, $\varepsilon_F=0$. In the range of $\Delta$ where $\delta$ is finite the chiral states would be the only remaining at $\varepsilon_F=0$, determining $G_{1D}=4e^2/h$ conductance of a device with electronic transport along a single DW. This range of parameters is painted in red on the inset in Fig. \ref{fig:5}(b). For the alternative range of parameters painted on the inset in blue, $\delta=0$, so that the additional non-chiral modes at $\varepsilon_F=0$ would increase the conductance value up to $G_{1D}=12e^2/h$ (where we account for the spin and valley degeneracy). 

Overall, the presented analysis of electronic properties of AB/BA boundary in bilayer graphene describes the parametric regimes where such a domain wall can host localised non-chiral 1D bands. We show that the existence of such bands is highly sensitive to the domain wall width and discuss their experimental manifestations.     

\section{Acknowledgments}
We thank J. Barrier, A. Garcia-Ruiz and P. Recher for usefull discussions. This work was supported by  EC-FET Core 3 European Graphene Flagship Project, EPSRC grants EP/S030719/1 and EP/V007033/1, and the Lloyd Register Foundation Nanotechnology Grant.
\bibliography{bibl}

\begin{thebibliography}{38}%
\makeatletter
\providecommand \@ifxundefined [1]{%
 \@ifx{#1\undefined}
}%
\providecommand \@ifnum [1]{%
 \ifnum #1\expandafter \@firstoftwo
 \else \expandafter \@secondoftwo
 \fi
}%
\providecommand \@ifx [1]{%
 \ifx #1\expandafter \@firstoftwo
 \else \expandafter \@secondoftwo
 \fi
}%
\providecommand \natexlab [1]{#1}%
\providecommand \enquote  [1]{``#1''}%
\providecommand \bibnamefont  [1]{#1}%
\providecommand \bibfnamefont [1]{#1}%
\providecommand \citenamefont [1]{#1}%
\providecommand \href@noop [0]{\@secondoftwo}%
\providecommand \href [0]{\begingroup \@sanitize@url \@href}%
\providecommand \@href[1]{\@@startlink{#1}\@@href}%
\providecommand \@@href[1]{\endgroup#1\@@endlink}%
\providecommand \@sanitize@url [0]{\catcode `\\12\catcode `\$12\catcode
  `\&12\catcode `\#12\catcode `\^12\catcode `\_12\catcode `\%12\relax}%
\providecommand \@@startlink[1]{}%
\providecommand \@@endlink[0]{}%
\providecommand \url  [0]{\begingroup\@sanitize@url \@url }%
\providecommand \@url [1]{\endgroup\@href {#1}{\urlprefix }}%
\providecommand \urlprefix  [0]{URL }%
\providecommand \Eprint [0]{\href }%
\providecommand \doibase [0]{https://doi.org/}%
\providecommand \selectlanguage [0]{\@gobble}%
\providecommand \bibinfo  [0]{\@secondoftwo}%
\providecommand \bibfield  [0]{\@secondoftwo}%
\providecommand \translation [1]{[#1]}%
\providecommand \BibitemOpen [0]{}%
\providecommand \bibitemStop [0]{}%
\providecommand \bibitemNoStop [0]{.\EOS\space}%
\providecommand \EOS [0]{\spacefactor3000\relax}%
\providecommand \BibitemShut  [1]{\csname bibitem#1\endcsname}%
\let\auto@bib@innerbib\@empty
\bibitem [{\citenamefont {Oostinga}\ \emph {et~al.}(2007)\citenamefont
  {Oostinga}, \citenamefont {Heersche}, \citenamefont {Liu}, \citenamefont
  {Morpurgo},\ and\ \citenamefont {Vandersypen}}]{Oostinga2007}%
  \BibitemOpen
  \bibfield  {author} {\bibinfo {author} {\bibfnamefont {J.~B.}\ \bibnamefont
  {Oostinga}}, \bibinfo {author} {\bibfnamefont {H.~B.}\ \bibnamefont
  {Heersche}}, \bibinfo {author} {\bibfnamefont {X.}~\bibnamefont {Liu}},
  \bibinfo {author} {\bibfnamefont {A.~F.}\ \bibnamefont {Morpurgo}},\ and\
  \bibinfo {author} {\bibfnamefont {L.~M.~K.}\ \bibnamefont {Vandersypen}},\
  }\bibfield  {title} {\bibinfo {title} {Gate-induced insulating state in
  bilayer graphene devices},\ }\href {https://doi.org/10.1038/nmat2082}
  {\bibfield  {journal} {\bibinfo  {journal} {Nature Materials}\ }\textbf
  {\bibinfo {volume} {7}},\ \bibinfo {pages} {151} (\bibinfo {year}
  {2007})}\BibitemShut {NoStop}%
\bibitem [{\citenamefont {Martin}\ \emph {et~al.}(2008)\citenamefont {Martin},
  \citenamefont {Blanter},\ and\ \citenamefont {Morpurgo}}]{MorpurgoPRL2008}%
  \BibitemOpen
  \bibfield  {author} {\bibinfo {author} {\bibfnamefont {I.}~\bibnamefont
  {Martin}}, \bibinfo {author} {\bibfnamefont {Y.~M.}\ \bibnamefont
  {Blanter}},\ and\ \bibinfo {author} {\bibfnamefont {A.~F.}\ \bibnamefont
  {Morpurgo}},\ }\bibfield  {title} {\bibinfo {title} {Topological confinement
  in bilayer graphene},\ }\href
  {https://doi.org/10.1103/PhysRevLett.100.036804} {\bibfield  {journal}
  {\bibinfo  {journal} {Phys. Rev. Lett.}\ }\textbf {\bibinfo {volume} {100}},\
  \bibinfo {pages} {036804} (\bibinfo {year} {2008})}\BibitemShut {NoStop}%
\bibitem [{\citenamefont {Zhang}\ \emph {et~al.}(2009)\citenamefont {Zhang},
  \citenamefont {Tang}, \citenamefont {Girit}, \citenamefont {Hao},
  \citenamefont {Martin}, \citenamefont {Zettl}, \citenamefont {Crommie},
  \citenamefont {Shen},\ and\ \citenamefont {Wang}}]{Zhang2009}%
  \BibitemOpen
  \bibfield  {author} {\bibinfo {author} {\bibfnamefont {Y.}~\bibnamefont
  {Zhang}}, \bibinfo {author} {\bibfnamefont {T.-T.}\ \bibnamefont {Tang}},
  \bibinfo {author} {\bibfnamefont {C.}~\bibnamefont {Girit}}, \bibinfo
  {author} {\bibfnamefont {Z.}~\bibnamefont {Hao}}, \bibinfo {author}
  {\bibfnamefont {M.~C.}\ \bibnamefont {Martin}}, \bibinfo {author}
  {\bibfnamefont {A.}~\bibnamefont {Zettl}}, \bibinfo {author} {\bibfnamefont
  {M.~F.}\ \bibnamefont {Crommie}}, \bibinfo {author} {\bibfnamefont {Y.~R.}\
  \bibnamefont {Shen}},\ and\ \bibinfo {author} {\bibfnamefont
  {F.}~\bibnamefont {Wang}},\ }\bibfield  {title} {\bibinfo {title} {Direct
  observation of a widely tunable bandgap in bilayer graphene},\ }\href
  {https://doi.org/10.1038/nature08105} {\bibfield  {journal} {\bibinfo
  {journal} {Nature}\ }\textbf {\bibinfo {volume} {459}},\ \bibinfo {pages}
  {820} (\bibinfo {year} {2009})}\BibitemShut {NoStop}%
\bibitem [{\citenamefont {Varlet}\ \emph {et~al.}(2014)\citenamefont {Varlet},
  \citenamefont {Liu}, \citenamefont {Krueckl}, \citenamefont {Bischoff},
  \citenamefont {Simonet}, \citenamefont {Watanabe}, \citenamefont {Taniguchi},
  \citenamefont {Richter}, \citenamefont {Ensslin},\ and\ \citenamefont
  {Ihn}}]{Varlet2014}%
  \BibitemOpen
  \bibfield  {author} {\bibinfo {author} {\bibfnamefont {A.}~\bibnamefont
  {Varlet}}, \bibinfo {author} {\bibfnamefont {M.-H.}\ \bibnamefont {Liu}},
  \bibinfo {author} {\bibfnamefont {V.}~\bibnamefont {Krueckl}}, \bibinfo
  {author} {\bibfnamefont {D.}~\bibnamefont {Bischoff}}, \bibinfo {author}
  {\bibfnamefont {P.}~\bibnamefont {Simonet}}, \bibinfo {author} {\bibfnamefont
  {K.}~\bibnamefont {Watanabe}}, \bibinfo {author} {\bibfnamefont
  {T.}~\bibnamefont {Taniguchi}}, \bibinfo {author} {\bibfnamefont
  {K.}~\bibnamefont {Richter}}, \bibinfo {author} {\bibfnamefont
  {K.}~\bibnamefont {Ensslin}},\ and\ \bibinfo {author} {\bibfnamefont
  {T.}~\bibnamefont {Ihn}},\ }\bibfield  {title} {\bibinfo {title}
  {Fabry-p{\'{e}}rot interference in gapped bilayer graphene with broken
  anti-klein tunneling},\ }\href
  {https://doi.org/10.1103/physrevlett.113.116601} {\bibfield  {journal}
  {\bibinfo  {journal} {Physical Review Letters}\ }\textbf {\bibinfo {volume}
  {113}},\ \bibinfo {pages} {116601} (\bibinfo {year} {2014})}\BibitemShut
  {NoStop}%
\bibitem [{\citenamefont {Cao}\ \emph {et~al.}(2018{\natexlab{a}})\citenamefont
  {Cao}, \citenamefont {Fatemi}, \citenamefont {Demir}, \citenamefont {Fang},
  \citenamefont {Tomarken}, \citenamefont {Luo}, \citenamefont
  {Sanchez-Yamagishi}, \citenamefont {Watanabe}, \citenamefont {Taniguchi},
  \citenamefont {Kaxiras},\ and\ \citenamefont {et~al}}]{cao2018correlated}%
  \BibitemOpen
  \bibfield  {author} {\bibinfo {author} {\bibfnamefont {Y.}~\bibnamefont
  {Cao}}, \bibinfo {author} {\bibfnamefont {V.}~\bibnamefont {Fatemi}},
  \bibinfo {author} {\bibfnamefont {A.}~\bibnamefont {Demir}}, \bibinfo
  {author} {\bibfnamefont {S.}~\bibnamefont {Fang}}, \bibinfo {author}
  {\bibfnamefont {S.~L.}\ \bibnamefont {Tomarken}}, \bibinfo {author}
  {\bibfnamefont {J.~Y.}\ \bibnamefont {Luo}}, \bibinfo {author} {\bibfnamefont
  {J.~D.}\ \bibnamefont {Sanchez-Yamagishi}}, \bibinfo {author} {\bibfnamefont
  {K.}~\bibnamefont {Watanabe}}, \bibinfo {author} {\bibfnamefont
  {T.}~\bibnamefont {Taniguchi}}, \bibinfo {author} {\bibfnamefont
  {E.}~\bibnamefont {Kaxiras}},\ and\ \bibinfo {author} {\bibnamefont
  {et~al}},\ }\bibfield  {title} {\bibinfo {title} {Correlated insulator
  behaviour at half-filling in magic-angle graphene superlattices},\ }\href
  {https://doi.org/10.1038/nature26154} {\bibfield  {journal} {\bibinfo
  {journal} {Nature}\ }\textbf {\bibinfo {volume} {556}},\ \bibinfo {pages}
  {80} (\bibinfo {year} {2018}{\natexlab{a}})}\BibitemShut {NoStop}%
\bibitem [{\citenamefont {Cao}\ \emph {et~al.}(2018{\natexlab{b}})\citenamefont
  {Cao}, \citenamefont {Fatemi}, \citenamefont {Fang}, \citenamefont
  {Watanabe}, \citenamefont {Taniguchi}, \citenamefont {Kaxiras},\ and\
  \citenamefont {Jarillo-Herrero}}]{cao2018unconventional}%
  \BibitemOpen
  \bibfield  {author} {\bibinfo {author} {\bibfnamefont {Y.}~\bibnamefont
  {Cao}}, \bibinfo {author} {\bibfnamefont {V.}~\bibnamefont {Fatemi}},
  \bibinfo {author} {\bibfnamefont {S.}~\bibnamefont {Fang}}, \bibinfo {author}
  {\bibfnamefont {K.}~\bibnamefont {Watanabe}}, \bibinfo {author}
  {\bibfnamefont {T.}~\bibnamefont {Taniguchi}}, \bibinfo {author}
  {\bibfnamefont {E.}~\bibnamefont {Kaxiras}},\ and\ \bibinfo {author}
  {\bibfnamefont {P.}~\bibnamefont {Jarillo-Herrero}},\ }\bibfield  {title}
  {\bibinfo {title} {Unconventional superconductivity in magic-angle graphene
  superlattices},\ }\href {https://doi.org/10.1038/nature26160} {\bibfield
  {journal} {\bibinfo  {journal} {Nature}\ }\textbf {\bibinfo {volume} {556}},\
  \bibinfo {pages} {43} (\bibinfo {year} {2018}{\natexlab{b}})}\BibitemShut
  {NoStop}%
\bibitem [{\citenamefont {Rickhaus}\ \emph {et~al.}(2018)\citenamefont
  {Rickhaus}, \citenamefont {Wallbank}, \citenamefont {Slizovskiy},
  \citenamefont {Pisoni}, \citenamefont {Overweg}, \citenamefont {Lee},
  \citenamefont {Eich}, \citenamefont {Liu}, \citenamefont {Watanabe},
  \citenamefont {Taniguchi}, \citenamefont {Ihn},\ and\ \citenamefont
  {Ensslin}}]{Rickhaus2018}%
  \BibitemOpen
  \bibfield  {author} {\bibinfo {author} {\bibfnamefont {P.}~\bibnamefont
  {Rickhaus}}, \bibinfo {author} {\bibfnamefont {J.}~\bibnamefont {Wallbank}},
  \bibinfo {author} {\bibfnamefont {S.}~\bibnamefont {Slizovskiy}}, \bibinfo
  {author} {\bibfnamefont {R.}~\bibnamefont {Pisoni}}, \bibinfo {author}
  {\bibfnamefont {H.}~\bibnamefont {Overweg}}, \bibinfo {author} {\bibfnamefont
  {Y.}~\bibnamefont {Lee}}, \bibinfo {author} {\bibfnamefont {M.}~\bibnamefont
  {Eich}}, \bibinfo {author} {\bibfnamefont {M.-H.}\ \bibnamefont {Liu}},
  \bibinfo {author} {\bibfnamefont {K.}~\bibnamefont {Watanabe}}, \bibinfo
  {author} {\bibfnamefont {T.}~\bibnamefont {Taniguchi}}, \bibinfo {author}
  {\bibfnamefont {T.}~\bibnamefont {Ihn}},\ and\ \bibinfo {author}
  {\bibfnamefont {K.}~\bibnamefont {Ensslin}},\ }\bibfield  {title} {\bibinfo
  {title} {Transport through a network of topological channels in twisted
  bilayer graphene},\ }\href {https://doi.org/10.1021/acs.nanolett.8b02387}
  {\bibfield  {journal} {\bibinfo  {journal} {Nano Letters}\ }\textbf {\bibinfo
  {volume} {18}},\ \bibinfo {pages} {6725} (\bibinfo {year}
  {2018})}\BibitemShut {NoStop}%
\bibitem [{\citenamefont {{Yankowitz}}\ \emph {et~al.}(2019)\citenamefont
  {{Yankowitz}}, \citenamefont {{Chen}}, \citenamefont {{Polshyn}},
  \citenamefont {{Zhang}}, \citenamefont {{Watanabe}}, \citenamefont
  {{Taniguchi}}, \citenamefont {{Graf}}, \citenamefont {{Young}},\ and\
  \citenamefont {{Dean}}}]{Yankowitz2019}%
  \BibitemOpen
  \bibfield  {author} {\bibinfo {author} {\bibfnamefont {M.}~\bibnamefont
  {{Yankowitz}}}, \bibinfo {author} {\bibfnamefont {S.}~\bibnamefont {{Chen}}},
  \bibinfo {author} {\bibfnamefont {H.}~\bibnamefont {{Polshyn}}}, \bibinfo
  {author} {\bibfnamefont {Y.}~\bibnamefont {{Zhang}}}, \bibinfo {author}
  {\bibfnamefont {K.}~\bibnamefont {{Watanabe}}}, \bibinfo {author}
  {\bibfnamefont {T.}~\bibnamefont {{Taniguchi}}}, \bibinfo {author}
  {\bibfnamefont {D.}~\bibnamefont {{Graf}}}, \bibinfo {author} {\bibfnamefont
  {A.~F.}\ \bibnamefont {{Young}}},\ and\ \bibinfo {author} {\bibfnamefont
  {C.~R.}\ \bibnamefont {{Dean}}},\ }\bibfield  {title} {\bibinfo {title}
  {{Tuning superconductivity in twisted bilayer graphene}},\ }\href
  {https://doi.org/10.1126/science.aav1910} {\bibfield  {journal} {\bibinfo
  {journal} {Science}\ }\textbf {\bibinfo {volume} {363}},\ \bibinfo {pages}
  {1059} (\bibinfo {year} {2019})}\BibitemShut {NoStop}%
\bibitem [{\citenamefont {Xu}\ \emph {et~al.}(2019)\citenamefont {Xu},
  \citenamefont {Berdyugin}, \citenamefont {Kumaravadivel}, \citenamefont
  {Guinea}, \citenamefont {Krishna~Kumar}, \citenamefont {Bandurin},
  \citenamefont {Morozov}, \citenamefont {Kuang}, \citenamefont {Tsim},
  \citenamefont {Liu}, \citenamefont {Edgar}, \citenamefont {Grigorieva},
  \citenamefont {Fal'ko}, \citenamefont {Kim},\ and\ \citenamefont
  {Geim}}]{Xu2019}%
  \BibitemOpen
  \bibfield  {author} {\bibinfo {author} {\bibfnamefont {S.~G.}\ \bibnamefont
  {Xu}}, \bibinfo {author} {\bibfnamefont {A.~I.}\ \bibnamefont {Berdyugin}},
  \bibinfo {author} {\bibfnamefont {P.}~\bibnamefont {Kumaravadivel}}, \bibinfo
  {author} {\bibfnamefont {F.}~\bibnamefont {Guinea}}, \bibinfo {author}
  {\bibfnamefont {R.}~\bibnamefont {Krishna~Kumar}}, \bibinfo {author}
  {\bibfnamefont {D.~A.}\ \bibnamefont {Bandurin}}, \bibinfo {author}
  {\bibfnamefont {S.~V.}\ \bibnamefont {Morozov}}, \bibinfo {author}
  {\bibfnamefont {W.}~\bibnamefont {Kuang}}, \bibinfo {author} {\bibfnamefont
  {B.}~\bibnamefont {Tsim}}, \bibinfo {author} {\bibfnamefont {S.}~\bibnamefont
  {Liu}}, \bibinfo {author} {\bibfnamefont {J.~H.}\ \bibnamefont {Edgar}},
  \bibinfo {author} {\bibfnamefont {I.~V.}\ \bibnamefont {Grigorieva}},
  \bibinfo {author} {\bibfnamefont {V.~I.}\ \bibnamefont {Fal'ko}}, \bibinfo
  {author} {\bibfnamefont {M.}~\bibnamefont {Kim}},\ and\ \bibinfo {author}
  {\bibfnamefont {A.~K.}\ \bibnamefont {Geim}},\ }\bibfield  {title} {\bibinfo
  {title} {Giant oscillations in a triangular network of one-dimensional states
  in marginally twisted graphene},\ }\href
  {https://doi.org/10.1038/s41467-019-11971-7} {\bibfield  {journal} {\bibinfo
  {journal} {Nature Communications}\ }\textbf {\bibinfo {volume} {10}},\
  \bibinfo {pages} {4008} (\bibinfo {year} {2019})}\BibitemShut {NoStop}%
\bibitem [{\citenamefont {Yoo}\ \emph {et~al.}(2019)\citenamefont {Yoo},
  \citenamefont {Engelke}, \citenamefont {Carr}, \citenamefont {Fang},
  \citenamefont {Zhang}, \citenamefont {Cazeaux}, \citenamefont {Sung},
  \citenamefont {Hovden}, \citenamefont {Tsen}, \citenamefont {Taniguchi},\
  and\ \citenamefont {et~al}}]{yoo2019atomic}%
  \BibitemOpen
  \bibfield  {author} {\bibinfo {author} {\bibfnamefont {H.}~\bibnamefont
  {Yoo}}, \bibinfo {author} {\bibfnamefont {R.}~\bibnamefont {Engelke}},
  \bibinfo {author} {\bibfnamefont {S.}~\bibnamefont {Carr}}, \bibinfo {author}
  {\bibfnamefont {S.}~\bibnamefont {Fang}}, \bibinfo {author} {\bibfnamefont
  {K.}~\bibnamefont {Zhang}}, \bibinfo {author} {\bibfnamefont
  {P.}~\bibnamefont {Cazeaux}}, \bibinfo {author} {\bibfnamefont {S.~H.}\
  \bibnamefont {Sung}}, \bibinfo {author} {\bibfnamefont {R.}~\bibnamefont
  {Hovden}}, \bibinfo {author} {\bibfnamefont {A.~W.}\ \bibnamefont {Tsen}},
  \bibinfo {author} {\bibfnamefont {T.}~\bibnamefont {Taniguchi}},\ and\
  \bibinfo {author} {\bibnamefont {et~al}},\ }\bibfield  {title} {\bibinfo
  {title} {Atomic and electronic reconstruction at the van der {W}aals
  interface in twisted bilayer graphene},\ }\href
  {https://doi.org/10.1038/s41563-019-0346-z} {\bibfield  {journal} {\bibinfo
  {journal} {Nature materials}\ }\textbf {\bibinfo {volume} {18}},\ \bibinfo
  {pages} {448} (\bibinfo {year} {2019})}\BibitemShut {NoStop}%
\bibitem [{\citenamefont {Lu}\ \emph {et~al.}(2019)\citenamefont {Lu},
  \citenamefont {Stepanov}, \citenamefont {Yang}, \citenamefont {Xie},
  \citenamefont {Aamir}, \citenamefont {Das}, \citenamefont {Urgell},
  \citenamefont {Watanabe}, \citenamefont {Taniguchi}, \citenamefont {Zhang}
  \emph {et~al.}}]{lu2019superconductors}%
  \BibitemOpen
  \bibfield  {author} {\bibinfo {author} {\bibfnamefont {X.}~\bibnamefont
  {Lu}}, \bibinfo {author} {\bibfnamefont {P.}~\bibnamefont {Stepanov}},
  \bibinfo {author} {\bibfnamefont {W.}~\bibnamefont {Yang}}, \bibinfo {author}
  {\bibfnamefont {M.}~\bibnamefont {Xie}}, \bibinfo {author} {\bibfnamefont
  {M.~A.}\ \bibnamefont {Aamir}}, \bibinfo {author} {\bibfnamefont
  {I.}~\bibnamefont {Das}}, \bibinfo {author} {\bibfnamefont {C.}~\bibnamefont
  {Urgell}}, \bibinfo {author} {\bibfnamefont {K.}~\bibnamefont {Watanabe}},
  \bibinfo {author} {\bibfnamefont {T.}~\bibnamefont {Taniguchi}}, \bibinfo
  {author} {\bibfnamefont {G.}~\bibnamefont {Zhang}}, \emph {et~al.},\
  }\bibfield  {title} {\bibinfo {title} {Superconductors, orbital magnets and
  correlated states in magic-angle bilayer graphene},\ }\href
  {https://doi.org/10.1038/s41586-019-1695-0} {\bibfield  {journal} {\bibinfo
  {journal} {Nature}\ }\textbf {\bibinfo {volume} {574}},\ \bibinfo {pages}
  {653} (\bibinfo {year} {2019})}\BibitemShut {NoStop}%
\bibitem [{\citenamefont {Sharpe}\ \emph {et~al.}(2019)\citenamefont {Sharpe},
  \citenamefont {Fox}, \citenamefont {Barnard}, \citenamefont {Finney},
  \citenamefont {Watanabe}, \citenamefont {Taniguchi}, \citenamefont
  {Kastner},\ and\ \citenamefont {Goldhaber-Gordon}}]{sharpe2019emergent}%
  \BibitemOpen
  \bibfield  {author} {\bibinfo {author} {\bibfnamefont {A.~L.}\ \bibnamefont
  {Sharpe}}, \bibinfo {author} {\bibfnamefont {E.~J.}\ \bibnamefont {Fox}},
  \bibinfo {author} {\bibfnamefont {A.~W.}\ \bibnamefont {Barnard}}, \bibinfo
  {author} {\bibfnamefont {J.}~\bibnamefont {Finney}}, \bibinfo {author}
  {\bibfnamefont {K.}~\bibnamefont {Watanabe}}, \bibinfo {author}
  {\bibfnamefont {T.}~\bibnamefont {Taniguchi}}, \bibinfo {author}
  {\bibfnamefont {M.}~\bibnamefont {Kastner}},\ and\ \bibinfo {author}
  {\bibfnamefont {D.}~\bibnamefont {Goldhaber-Gordon}},\ }\bibfield  {title}
  {\bibinfo {title} {Emergent ferromagnetism near three-quarters filling in
  twisted bilayer graphene},\ }\href {https://doi.org/10.1126/science.aaw3780}
  {\bibfield  {journal} {\bibinfo  {journal} {Science}\ }\textbf {\bibinfo
  {volume} {365}},\ \bibinfo {pages} {605} (\bibinfo {year}
  {2019})}\BibitemShut {NoStop}%
\bibitem [{\citenamefont {Cao}\ \emph {et~al.}(2021)\citenamefont {Cao},
  \citenamefont {Rodan-Legrain}, \citenamefont {Park}, \citenamefont {Yuan},
  \citenamefont {Watanabe}, \citenamefont {Taniguchi}, \citenamefont
  {Fernandes}, \citenamefont {Fu},\ and\ \citenamefont
  {Jarillo-Herrero}}]{cao2021nematicity}%
  \BibitemOpen
  \bibfield  {author} {\bibinfo {author} {\bibfnamefont {Y.}~\bibnamefont
  {Cao}}, \bibinfo {author} {\bibfnamefont {D.}~\bibnamefont {Rodan-Legrain}},
  \bibinfo {author} {\bibfnamefont {J.~M.}\ \bibnamefont {Park}}, \bibinfo
  {author} {\bibfnamefont {N.~F.}\ \bibnamefont {Yuan}}, \bibinfo {author}
  {\bibfnamefont {K.}~\bibnamefont {Watanabe}}, \bibinfo {author}
  {\bibfnamefont {T.}~\bibnamefont {Taniguchi}}, \bibinfo {author}
  {\bibfnamefont {R.~M.}\ \bibnamefont {Fernandes}}, \bibinfo {author}
  {\bibfnamefont {L.}~\bibnamefont {Fu}},\ and\ \bibinfo {author}
  {\bibfnamefont {P.}~\bibnamefont {Jarillo-Herrero}},\ }\bibfield  {title}
  {\bibinfo {title} {Nematicity and competing orders in superconducting
  magic-angle graphene},\ }\href {https://doi.org/10.1126/science.abc2836}
  {\bibfield  {journal} {\bibinfo  {journal} {Science}\ }\textbf {\bibinfo
  {volume} {372}},\ \bibinfo {pages} {264} (\bibinfo {year}
  {2021})}\BibitemShut {NoStop}%
\bibitem [{\citenamefont {McCann}\ and\ \citenamefont
  {Fal'ko}(2006)}]{PhysRevLett.96.086805}%
  \BibitemOpen
  \bibfield  {author} {\bibinfo {author} {\bibfnamefont {E.}~\bibnamefont
  {McCann}}\ and\ \bibinfo {author} {\bibfnamefont {V.~I.}\ \bibnamefont
  {Fal'ko}},\ }\bibfield  {title} {\bibinfo {title} {Landau-level degeneracy
  and quantum hall effect in a graphite bilayer},\ }\href
  {https://doi.org/10.1103/PhysRevLett.96.086805} {\bibfield  {journal}
  {\bibinfo  {journal} {Phys. Rev. Lett.}\ }\textbf {\bibinfo {volume} {96}},\
  \bibinfo {pages} {086805} (\bibinfo {year} {2006})}\BibitemShut {NoStop}%
\bibitem [{\citenamefont {McCann}(2006)}]{McCann2006}%
  \BibitemOpen
  \bibfield  {author} {\bibinfo {author} {\bibfnamefont {E.}~\bibnamefont
  {McCann}},\ }\bibfield  {title} {\bibinfo {title} {Asymmetry gap in the
  electronic band structure of bilayer graphene},\ }\href
  {https://doi.org/10.1103/physrevb.74.161403} {\bibfield  {journal} {\bibinfo
  {journal} {Physical Review B}\ }\textbf {\bibinfo {volume} {74}},\ \bibinfo
  {pages} {161403} (\bibinfo {year} {2006})}\BibitemShut {NoStop}%
\bibitem [{\citenamefont {Slizovskiy}\ \emph {et~al.}(2021)\citenamefont
  {Slizovskiy}, \citenamefont {Garcia-Ruiz}, \citenamefont {Berdyugin},
  \citenamefont {Xin}, \citenamefont {Taniguchi}, \citenamefont {Watanabe},
  \citenamefont {Geim}, \citenamefont {Drummond},\ and\ \citenamefont
  {Fal'ko}}]{Slizovskiy2021}%
  \BibitemOpen
  \bibfield  {author} {\bibinfo {author} {\bibfnamefont {S.}~\bibnamefont
  {Slizovskiy}}, \bibinfo {author} {\bibfnamefont {A.}~\bibnamefont
  {Garcia-Ruiz}}, \bibinfo {author} {\bibfnamefont {A.~I.}\ \bibnamefont
  {Berdyugin}}, \bibinfo {author} {\bibfnamefont {N.}~\bibnamefont {Xin}},
  \bibinfo {author} {\bibfnamefont {T.}~\bibnamefont {Taniguchi}}, \bibinfo
  {author} {\bibfnamefont {K.}~\bibnamefont {Watanabe}}, \bibinfo {author}
  {\bibfnamefont {A.~K.}\ \bibnamefont {Geim}}, \bibinfo {author}
  {\bibfnamefont {N.~D.}\ \bibnamefont {Drummond}},\ and\ \bibinfo {author}
  {\bibfnamefont {V.~I.}\ \bibnamefont {Fal'ko}},\ }\bibfield  {title}
  {\bibinfo {title} {Out-of-plane dielectric susceptibility of graphene in
  twistronic and bernal bilayers},\ }\href
  {https://doi.org/10.1021/acs.nanolett.1c02211} {\bibfield  {journal}
  {\bibinfo  {journal} {Nano Letters}\ }\textbf {\bibinfo {volume} {21}},\
  \bibinfo {pages} {6678} (\bibinfo {year} {2021})}\BibitemShut {NoStop}%
\bibitem [{\citenamefont {Zhang}\ \emph {et~al.}(2013)\citenamefont {Zhang},
  \citenamefont {MacDonald},\ and\ \citenamefont {Mele}}]{ZhangMacDMele2013}%
  \BibitemOpen
  \bibfield  {author} {\bibinfo {author} {\bibfnamefont {F.}~\bibnamefont
  {Zhang}}, \bibinfo {author} {\bibfnamefont {A.~H.}\ \bibnamefont
  {MacDonald}},\ and\ \bibinfo {author} {\bibfnamefont {E.~J.}\ \bibnamefont
  {Mele}},\ }\bibfield  {title} {\bibinfo {title} {Valley chern numbers and
  boundary modes in gapped bilayer graphene},\ }\href
  {https://doi.org/10.1073/pnas.1308853110} {\bibfield  {journal} {\bibinfo
  {journal} {Proceedings of the National Academy of Sciences}\ }\textbf
  {\bibinfo {volume} {110}},\ \bibinfo {pages} {10546} (\bibinfo {year}
  {2013})}\BibitemShut {NoStop}%
\bibitem [{\citenamefont {Delavignette}\ and\ \citenamefont
  {Amelinckx}(1960)}]{Delavignette1960}%
  \BibitemOpen
  \bibfield  {author} {\bibinfo {author} {\bibfnamefont {P.}~\bibnamefont
  {Delavignette}}\ and\ \bibinfo {author} {\bibfnamefont {S.}~\bibnamefont
  {Amelinckx}},\ }\bibfield  {title} {\bibinfo {title} {Dislocation ribbons and
  stacking faults in graphite},\ }\href {https://doi.org/10.1063/1.1735924}
  {\bibfield  {journal} {\bibinfo  {journal} {Journal of Applied Physics}\
  }\textbf {\bibinfo {volume} {31}},\ \bibinfo {pages} {1691} (\bibinfo {year}
  {1960})}\BibitemShut {NoStop}%
\bibitem [{\citenamefont {Butz}\ \emph {et~al.}(2013)\citenamefont {Butz},
  \citenamefont {Dolle}, \citenamefont {Niekiel}, \citenamefont {Weber},
  \citenamefont {Waldmann}, \citenamefont {Weber}, \citenamefont {Meyer},\ and\
  \citenamefont {Spiecker}}]{Butz2013}%
  \BibitemOpen
  \bibfield  {author} {\bibinfo {author} {\bibfnamefont {B.}~\bibnamefont
  {Butz}}, \bibinfo {author} {\bibfnamefont {C.}~\bibnamefont {Dolle}},
  \bibinfo {author} {\bibfnamefont {F.}~\bibnamefont {Niekiel}}, \bibinfo
  {author} {\bibfnamefont {K.}~\bibnamefont {Weber}}, \bibinfo {author}
  {\bibfnamefont {D.}~\bibnamefont {Waldmann}}, \bibinfo {author}
  {\bibfnamefont {H.~B.}\ \bibnamefont {Weber}}, \bibinfo {author}
  {\bibfnamefont {B.}~\bibnamefont {Meyer}},\ and\ \bibinfo {author}
  {\bibfnamefont {E.}~\bibnamefont {Spiecker}},\ }\bibfield  {title} {\bibinfo
  {title} {Dislocations in bilayer graphene},\ }\href
  {https://doi.org/10.1038/nature12780} {\bibfield  {journal} {\bibinfo
  {journal} {Nature}\ }\textbf {\bibinfo {volume} {505}},\ \bibinfo {pages}
  {533} (\bibinfo {year} {2013})}\BibitemShut {NoStop}%
\bibitem [{\citenamefont {Alden}\ \emph {et~al.}(2013)\citenamefont {Alden},
  \citenamefont {Tsen}, \citenamefont {Huang}, \citenamefont {Hovden},
  \citenamefont {Brown}, \citenamefont {Park}, \citenamefont {Muller},\ and\
  \citenamefont {McEuen}}]{AldenPNAS}%
  \BibitemOpen
  \bibfield  {author} {\bibinfo {author} {\bibfnamefont {J.~S.}\ \bibnamefont
  {Alden}}, \bibinfo {author} {\bibfnamefont {A.~W.}\ \bibnamefont {Tsen}},
  \bibinfo {author} {\bibfnamefont {P.~Y.}\ \bibnamefont {Huang}}, \bibinfo
  {author} {\bibfnamefont {R.}~\bibnamefont {Hovden}}, \bibinfo {author}
  {\bibfnamefont {L.}~\bibnamefont {Brown}}, \bibinfo {author} {\bibfnamefont
  {J.}~\bibnamefont {Park}}, \bibinfo {author} {\bibfnamefont {D.~A.}\
  \bibnamefont {Muller}},\ and\ \bibinfo {author} {\bibfnamefont {P.~L.}\
  \bibnamefont {McEuen}},\ }\bibfield  {title} {\bibinfo {title} {Strain
  solitons and topological defects in bilayer graphene},\ }\href
  {https://doi.org/10.1073/pnas.1309394110} {\bibfield  {journal} {\bibinfo
  {journal} {PNAS}\ }\textbf {\bibinfo {volume} {110}},\ \bibinfo {pages}
  {11256} (\bibinfo {year} {2013})}\BibitemShut {NoStop}%
\bibitem [{\citenamefont {Hirth}\ and\ \citenamefont
  {Lothe}(1992)}]{hirth1992}%
  \BibitemOpen
  \bibfield  {author} {\bibinfo {author} {\bibfnamefont {J.~P.}\ \bibnamefont
  {Hirth}}\ and\ \bibinfo {author} {\bibfnamefont {J.}~\bibnamefont {Lothe}},\
  }\href@noop {} {\emph {\bibinfo {title} {Theory of Dislocations}}}\ (\bibinfo
   {publisher} {Krieger Publishing Company},\ \bibinfo {year}
  {1992})\BibitemShut {NoStop}%
\bibitem [{\citenamefont {Kosevich}(2005)}]{Kosevich2005}%
  \BibitemOpen
  \bibfield  {author} {\bibinfo {author} {\bibfnamefont {A.~M.}\ \bibnamefont
  {Kosevich}},\ }\href {https://doi.org/10.1002/352760667x} {\emph {\bibinfo
  {title} {The Crystal Lattice}}}\ (\bibinfo  {publisher} {Wiley},\ \bibinfo
  {year} {2005})\BibitemShut {NoStop}%
\bibitem [{\citenamefont {Ju}\ \emph {et~al.}(2015)\citenamefont {Ju},
  \citenamefont {Shi}, \citenamefont {Nair}, \citenamefont {Lv}, \citenamefont
  {Jin}, \citenamefont {Velasco}, \citenamefont {Ojeda-Aristizabal},
  \citenamefont {Bechtel}, \citenamefont {Martin}, \citenamefont {Zettl},
  \citenamefont {Analytis},\ and\ \citenamefont {Wang}}]{Ju2015}%
  \BibitemOpen
  \bibfield  {author} {\bibinfo {author} {\bibfnamefont {L.}~\bibnamefont
  {Ju}}, \bibinfo {author} {\bibfnamefont {Z.}~\bibnamefont {Shi}}, \bibinfo
  {author} {\bibfnamefont {N.}~\bibnamefont {Nair}}, \bibinfo {author}
  {\bibfnamefont {Y.}~\bibnamefont {Lv}}, \bibinfo {author} {\bibfnamefont
  {C.}~\bibnamefont {Jin}}, \bibinfo {author} {\bibfnamefont {J.}~\bibnamefont
  {Velasco}}, \bibinfo {author} {\bibfnamefont {C.}~\bibnamefont
  {Ojeda-Aristizabal}}, \bibinfo {author} {\bibfnamefont {H.~A.}\ \bibnamefont
  {Bechtel}}, \bibinfo {author} {\bibfnamefont {M.~C.}\ \bibnamefont {Martin}},
  \bibinfo {author} {\bibfnamefont {A.}~\bibnamefont {Zettl}}, \bibinfo
  {author} {\bibfnamefont {J.}~\bibnamefont {Analytis}},\ and\ \bibinfo
  {author} {\bibfnamefont {F.}~\bibnamefont {Wang}},\ }\bibfield  {title}
  {\bibinfo {title} {Topological valley transport at bilayer graphene domain
  walls},\ }\href {https://doi.org/10.1038/nature14364} {\bibfield  {journal}
  {\bibinfo  {journal} {Nature}\ }\textbf {\bibinfo {volume} {520}},\ \bibinfo
  {pages} {650} (\bibinfo {year} {2015})}\BibitemShut {NoStop}%
\bibitem [{\citenamefont {Yin}\ \emph {et~al.}(2016)\citenamefont {Yin},
  \citenamefont {Jiang}, \citenamefont {Qiao},\ and\ \citenamefont
  {He}}]{Yin2016}%
  \BibitemOpen
  \bibfield  {author} {\bibinfo {author} {\bibfnamefont {L.-J.}\ \bibnamefont
  {Yin}}, \bibinfo {author} {\bibfnamefont {H.}~\bibnamefont {Jiang}}, \bibinfo
  {author} {\bibfnamefont {J.-B.}\ \bibnamefont {Qiao}},\ and\ \bibinfo
  {author} {\bibfnamefont {L.}~\bibnamefont {He}},\ }\bibfield  {title}
  {\bibinfo {title} {Direct imaging of topological edge states at a bilayer
  graphene domain wall},\ }\href {https://doi.org/10.1038/ncomms11760}
  {\bibfield  {journal} {\bibinfo  {journal} {Nature Communications}\ }\textbf
  {\bibinfo {volume} {7}},\ \bibinfo {pages} {11760} (\bibinfo {year}
  {2016})}\BibitemShut {NoStop}%
\bibitem [{\citenamefont {San-Jose}\ and\ \citenamefont
  {Prada}(2013)}]{SanJose2013}%
  \BibitemOpen
  \bibfield  {author} {\bibinfo {author} {\bibfnamefont {P.}~\bibnamefont
  {San-Jose}}\ and\ \bibinfo {author} {\bibfnamefont {E.}~\bibnamefont
  {Prada}},\ }\bibfield  {title} {\bibinfo {title} {Helical networks in twisted
  bilayer graphene under interlayer bias},\ }\href
  {https://doi.org/10.1103/physrevb.88.121408} {\bibfield  {journal} {\bibinfo
  {journal} {Physical Review B}\ }\textbf {\bibinfo {volume} {88}},\ \bibinfo
  {pages} {121408} (\bibinfo {year} {2013})}\BibitemShut {NoStop}%
\bibitem [{\citenamefont {Efimkin}\ and\ \citenamefont
  {MacDonald}(2018)}]{Efimkin2018}%
  \BibitemOpen
  \bibfield  {author} {\bibinfo {author} {\bibfnamefont {D.~K.}\ \bibnamefont
  {Efimkin}}\ and\ \bibinfo {author} {\bibfnamefont {A.~H.}\ \bibnamefont
  {MacDonald}},\ }\bibfield  {title} {\bibinfo {title} {Helical network model
  for twisted bilayer graphene},\ }\href
  {https://doi.org/10.1103/PhysRevB.98.035404} {\bibfield  {journal} {\bibinfo
  {journal} {Phys. Rev. B}\ }\textbf {\bibinfo {volume} {98}},\ \bibinfo
  {pages} {035404} (\bibinfo {year} {2018})}\BibitemShut {NoStop}%
\bibitem [{\citenamefont {Huang}\ \emph {et~al.}(2018)\citenamefont {Huang},
  \citenamefont {Kim}, \citenamefont {Efimkin}, \citenamefont {Lovorn},
  \citenamefont {Taniguchi}, \citenamefont {Watanabe}, \citenamefont
  {MacDonald}, \citenamefont {Tutuc},\ and\ \citenamefont {LeRoy}}]{Huang2018}%
  \BibitemOpen
  \bibfield  {author} {\bibinfo {author} {\bibfnamefont {S.}~\bibnamefont
  {Huang}}, \bibinfo {author} {\bibfnamefont {K.}~\bibnamefont {Kim}}, \bibinfo
  {author} {\bibfnamefont {D.~K.}\ \bibnamefont {Efimkin}}, \bibinfo {author}
  {\bibfnamefont {T.}~\bibnamefont {Lovorn}}, \bibinfo {author} {\bibfnamefont
  {T.}~\bibnamefont {Taniguchi}}, \bibinfo {author} {\bibfnamefont
  {K.}~\bibnamefont {Watanabe}}, \bibinfo {author} {\bibfnamefont {A.~H.}\
  \bibnamefont {MacDonald}}, \bibinfo {author} {\bibfnamefont {E.}~\bibnamefont
  {Tutuc}},\ and\ \bibinfo {author} {\bibfnamefont {B.~J.}\ \bibnamefont
  {LeRoy}},\ }\bibfield  {title} {\bibinfo {title} {Topologically protected
  helical states in minimally twisted bilayer graphene},\ }\href
  {https://doi.org/10.1103/physrevlett.121.037702} {\bibfield  {journal}
  {\bibinfo  {journal} {Physical Review Letters}\ }\textbf {\bibinfo {volume}
  {121}},\ \bibinfo {pages} {037702} (\bibinfo {year} {2018})}\BibitemShut
  {NoStop}%
\bibitem [{\citenamefont {Beule}\ \emph {et~al.}(2021)\citenamefont {Beule},
  \citenamefont {Dominguez},\ and\ \citenamefont {Recher}}]{Beule2021}%
  \BibitemOpen
  \bibfield  {author} {\bibinfo {author} {\bibfnamefont {C.~D.}\ \bibnamefont
  {Beule}}, \bibinfo {author} {\bibfnamefont {F.}~\bibnamefont {Dominguez}},\
  and\ \bibinfo {author} {\bibfnamefont {P.}~\bibnamefont {Recher}},\
  }\bibfield  {title} {\bibinfo {title} {Network model and four-terminal
  transport in minimally twisted bilayer graphene},\ }\href
  {https://doi.org/10.1103/physrevb.104.195410} {\bibfield  {journal} {\bibinfo
   {journal} {Physical Review B}\ }\textbf {\bibinfo {volume} {104}},\ \bibinfo
  {pages} {195410} (\bibinfo {year} {2021})}\BibitemShut {NoStop}%
\bibitem [{\citenamefont {Lane}\ \emph {et~al.}(2018)\citenamefont {Lane},
  \citenamefont {Andjelkovi{\'{c}}}, \citenamefont {Wallbank}, \citenamefont
  {Covaci}, \citenamefont {Peeters},\ and\ \citenamefont {Fal'ko}}]{Lane2018}%
  \BibitemOpen
  \bibfield  {author} {\bibinfo {author} {\bibfnamefont {T.~L.~M.}\
  \bibnamefont {Lane}}, \bibinfo {author} {\bibfnamefont {M.}~\bibnamefont
  {Andjelkovi{\'{c}}}}, \bibinfo {author} {\bibfnamefont {J.~R.}\ \bibnamefont
  {Wallbank}}, \bibinfo {author} {\bibfnamefont {L.}~\bibnamefont {Covaci}},
  \bibinfo {author} {\bibfnamefont {F.~M.}\ \bibnamefont {Peeters}},\ and\
  \bibinfo {author} {\bibfnamefont {V.~I.}\ \bibnamefont {Fal'ko}},\ }\bibfield
   {title} {\bibinfo {title} {Ballistic electron channels including weakly
  protected topological states in delaminated bilayer graphene},\ }\href
  {https://doi.org/10.1103/physrevb.97.045301} {\bibfield  {journal} {\bibinfo
  {journal} {Physical Review B}\ }\textbf {\bibinfo {volume} {97}},\ \bibinfo
  {pages} {045301} (\bibinfo {year} {2018})}\BibitemShut {NoStop}%
\bibitem [{\citenamefont {Theil}\ \emph {et~al.}(2022)\citenamefont {Theil},
  \citenamefont {Gupta}, \citenamefont {Wullschlaeger}, \citenamefont {Meyer},
  \citenamefont {Sharma},\ and\ \citenamefont {Shallcross}}]{Shallcross2022}%
  \BibitemOpen
  \bibfield  {author} {\bibinfo {author} {\bibfnamefont {S.}~\bibnamefont
  {Theil}}, \bibinfo {author} {\bibfnamefont {R.}~\bibnamefont {Gupta}},
  \bibinfo {author} {\bibfnamefont {F.}~\bibnamefont {Wullschlaeger}}, \bibinfo
  {author} {\bibfnamefont {B.}~\bibnamefont {Meyer}}, \bibinfo {author}
  {\bibfnamefont {S.}~\bibnamefont {Sharma}},\ and\ \bibinfo {author}
  {\bibfnamefont {S.}~\bibnamefont {Shallcross}},\ }\bibfield  {title}
  {\bibinfo {title} {Volkov-pankratov states in a 2d material: excited states
  of a structural soliton},\ }\href@noop {} {\bibfield  {journal} {\bibinfo
  {journal} {arXiv preprint arXiv:2203.16079}\ } (\bibinfo {year}
  {2022})}\BibitemShut {NoStop}%
\bibitem [{\citenamefont {Zhou}\ \emph {et~al.}(2015)\citenamefont {Zhou},
  \citenamefont {Han}, \citenamefont {Dai}, \citenamefont {Sun},\ and\
  \citenamefont {Srolovitz}}]{Zhou2015}%
  \BibitemOpen
  \bibfield  {author} {\bibinfo {author} {\bibfnamefont {S.}~\bibnamefont
  {Zhou}}, \bibinfo {author} {\bibfnamefont {J.}~\bibnamefont {Han}}, \bibinfo
  {author} {\bibfnamefont {S.}~\bibnamefont {Dai}}, \bibinfo {author}
  {\bibfnamefont {J.}~\bibnamefont {Sun}},\ and\ \bibinfo {author}
  {\bibfnamefont {D.~J.}\ \bibnamefont {Srolovitz}},\ }\bibfield  {title}
  {\bibinfo {title} {van der waals bilayer energetics: Generalized
  stacking-fault energy of graphene, boron nitride, and graphene/boron nitride
  bilayers},\ }\href {https://doi.org/10.1103/physrevb.92.155438} {\bibfield
  {journal} {\bibinfo  {journal} {Physical Review B}\ }\textbf {\bibinfo
  {volume} {92}},\ \bibinfo {pages} {155438} (\bibinfo {year}
  {2015})}\BibitemShut {NoStop}%
\bibitem [{\citenamefont {Enaldiev}\ \emph {et~al.}(2020)\citenamefont
  {Enaldiev}, \citenamefont {Z\'olyomi}, \citenamefont {Yelgel}, \citenamefont
  {Magorrian},\ and\ \citenamefont {Fal'ko}}]{Enaldiev_PRL}%
  \BibitemOpen
  \bibfield  {author} {\bibinfo {author} {\bibfnamefont {V.~V.}\ \bibnamefont
  {Enaldiev}}, \bibinfo {author} {\bibfnamefont {V.}~\bibnamefont {Z\'olyomi}},
  \bibinfo {author} {\bibfnamefont {C.}~\bibnamefont {Yelgel}}, \bibinfo
  {author} {\bibfnamefont {S.~J.}\ \bibnamefont {Magorrian}},\ and\ \bibinfo
  {author} {\bibfnamefont {V.~I.}\ \bibnamefont {Fal'ko}},\ }\bibfield  {title}
  {\bibinfo {title} {Stacking domains and dislocation networks in marginally
  twisted bilayers of transition metal dichalcogenides},\ }\href
  {https://doi.org/10.1103/PhysRevLett.124.206101} {\bibfield  {journal}
  {\bibinfo  {journal} {Phys. Rev. Lett.}\ }\textbf {\bibinfo {volume} {124}},\
  \bibinfo {pages} {206101} (\bibinfo {year} {2020})}\BibitemShut {NoStop}%
\bibitem [{\citenamefont {Popov}\ \emph {et~al.}(2011)\citenamefont {Popov},
  \citenamefont {Lebedeva}, \citenamefont {Knizhnik}, \citenamefont {Lozovik},\
  and\ \citenamefont {Potapkin}}]{Popov2011}%
  \BibitemOpen
  \bibfield  {author} {\bibinfo {author} {\bibfnamefont {A.~M.}\ \bibnamefont
  {Popov}}, \bibinfo {author} {\bibfnamefont {I.~V.}\ \bibnamefont {Lebedeva}},
  \bibinfo {author} {\bibfnamefont {A.~A.}\ \bibnamefont {Knizhnik}}, \bibinfo
  {author} {\bibfnamefont {Y.~E.}\ \bibnamefont {Lozovik}},\ and\ \bibinfo
  {author} {\bibfnamefont {B.~V.}\ \bibnamefont {Potapkin}},\ }\bibfield
  {title} {\bibinfo {title} {Commensurate-incommensurate phase transition in
  bilayer graphene},\ }\href {https://doi.org/10.1103/physrevb.84.045404}
  {\bibfield  {journal} {\bibinfo  {journal} {Physical Review B}\ }\textbf
  {\bibinfo {volume} {84}},\ \bibinfo {pages} {045404} (\bibinfo {year}
  {2011})}\BibitemShut {NoStop}%
\bibitem [{\citenamefont {Lebedeva}\ \emph {et~al.}(2016)\citenamefont
  {Lebedeva}, \citenamefont {Lebedev}, \citenamefont {Popov},\ and\
  \citenamefont {Knizhnik}}]{Lebedeva2016}%
  \BibitemOpen
  \bibfield  {author} {\bibinfo {author} {\bibfnamefont {I.~V.}\ \bibnamefont
  {Lebedeva}}, \bibinfo {author} {\bibfnamefont {A.~V.}\ \bibnamefont
  {Lebedev}}, \bibinfo {author} {\bibfnamefont {A.~M.}\ \bibnamefont {Popov}},\
  and\ \bibinfo {author} {\bibfnamefont {A.~A.}\ \bibnamefont {Knizhnik}},\
  }\bibfield  {title} {\bibinfo {title} {Dislocations in stacking and
  commensurate-incommensurate phase transition in bilayer graphene and
  hexagonal boron nitride},\ }\href
  {https://doi.org/10.1103/physrevb.93.235414} {\bibfield  {journal} {\bibinfo
  {journal} {Physical Review B}\ }\textbf {\bibinfo {volume} {93}},\ \bibinfo
  {pages} {235414} (\bibinfo {year} {2016})}\BibitemShut {NoStop}%
\bibitem [{\citenamefont {Lebedeva}\ and\ \citenamefont
  {Popov}(2019)}]{Lebedeva2019}%
  \BibitemOpen
  \bibfield  {author} {\bibinfo {author} {\bibfnamefont {I.~V.}\ \bibnamefont
  {Lebedeva}}\ and\ \bibinfo {author} {\bibfnamefont {A.~M.}\ \bibnamefont
  {Popov}},\ }\bibfield  {title} {\bibinfo {title} {Commensurate-incommensurate
  phase transition and a network of domain walls in bilayer graphene with a
  biaxially stretched layer},\ }\href
  {https://doi.org/10.1103/physrevb.99.195448} {\bibfield  {journal} {\bibinfo
  {journal} {Physical Review B}\ }\textbf {\bibinfo {volume} {99}},\ \bibinfo
  {pages} {195448} (\bibinfo {year} {2019})}\BibitemShut {NoStop}%
\bibitem [{\citenamefont {Carr}\ \emph {et~al.}(2018)\citenamefont {Carr},
  \citenamefont {Massatt}, \citenamefont {Torrisi}, \citenamefont {Cazeaux},
  \citenamefont {Luskin},\ and\ \citenamefont {Kaxiras}}]{CarrPRB2018}%
  \BibitemOpen
  \bibfield  {author} {\bibinfo {author} {\bibfnamefont {S.}~\bibnamefont
  {Carr}}, \bibinfo {author} {\bibfnamefont {D.}~\bibnamefont {Massatt}},
  \bibinfo {author} {\bibfnamefont {S.~B.}\ \bibnamefont {Torrisi}}, \bibinfo
  {author} {\bibfnamefont {P.}~\bibnamefont {Cazeaux}}, \bibinfo {author}
  {\bibfnamefont {M.}~\bibnamefont {Luskin}},\ and\ \bibinfo {author}
  {\bibfnamefont {E.}~\bibnamefont {Kaxiras}},\ }\bibfield  {title} {\bibinfo
  {title} {Relaxation and domain formation in incommensurate two-dimensional
  heterostructures},\ }\href {https://doi.org/10.1103/PhysRevB.98.224102}
  {\bibfield  {journal} {\bibinfo  {journal} {Phys. Rev. B}\ }\textbf {\bibinfo
  {volume} {98}},\ \bibinfo {pages} {224102} (\bibinfo {year}
  {2018})}\BibitemShut {NoStop}%
\bibitem [{Note1()}]{Note1}%
  \BibitemOpen
  \bibinfo {note} {Substituting in Eq. \protect \textup {\hbox {\mathsurround
  \z@ \protect \normalfont (\ignorespaces \ref {Eq:dw}\unskip \@@italiccorr )}}
  $a=0.246$\protect \,nm, $\mu = 920$\protect \,eV/nm$^2$\protect \, and
  $V=0.0296$\protect \,eV/nm$^2$ \cite {CarrPRB2018,Zhou2015} one obtains
  $w\approx 6$\protect \,nm.}\BibitemShut {Stop}%
\bibitem [{\citenamefont {Garcia-Ruiz}\ \emph {et~al.}(2021)\citenamefont
  {Garcia-Ruiz}, \citenamefont {Deng}, \citenamefont {Enaldiev},\ and\
  \citenamefont {Fal'ko}}]{PhysRevB.104.085402}%
  \BibitemOpen
  \bibfield  {author} {\bibinfo {author} {\bibfnamefont {A.}~\bibnamefont
  {Garcia-Ruiz}}, \bibinfo {author} {\bibfnamefont {H.-Y.}\ \bibnamefont
  {Deng}}, \bibinfo {author} {\bibfnamefont {V.~V.}\ \bibnamefont {Enaldiev}},\
  and\ \bibinfo {author} {\bibfnamefont {V.~I.}\ \bibnamefont {Fal'ko}},\
  }\bibfield  {title} {\bibinfo {title} {Full slonczewski-weiss-mcclure
  parametrization of few-layer twistronic graphene},\ }\href
  {https://doi.org/10.1103/PhysRevB.104.085402} {\bibfield  {journal} {\bibinfo
   {journal} {Phys. Rev. B}\ }\textbf {\bibinfo {volume} {104}},\ \bibinfo
  {pages} {085402} (\bibinfo {year} {2021})}\BibitemShut {NoStop}%
\end{thebibliography}%

\end{document}